# Controlled and Uncontrolled Stochastic Norton-Simon-Massagué Tumor Growth Models

Zehor Belkhatir[1], Michele Pavon[2], James C. Mathews[1], Maryam Pouryahya[1],

Joseph O. Deasy[1], Larry Norton[3], and Allen R. Tannenbaum[4]

## Abstract

Tumorigenesis is a complex process that is heterogeneous and affected by numerous sources of variability. This study presents a stochastic extension of a biologically grounded tumor growth model, referred to as the Norton-Simon-Massagué (NSM) tumor growth model. We first study the uncontrolled version of the model where the effect of chemotherapeutic drug agent is absent. Conditions on the model's parameters are derived to guarantee the positivity of the tumor volume and hence the validity of the proposed stochastic NSM model. To calibrate the proposed model we utilize a maximum likelihood-based estimation algorithm and population mixed-effect modeling formulation. The algorithm is tested by fitting previously published tumor volume mice data. Then, we study the controlled version of the model which includes the effect of chemotherapy treatment. Analysis of the influence of adding the control drug agent into the model and how sensitive it is to the stochastic parameters is performed both in open-loop and closed-loop viewpoints through different numerical simulations.

## I. INTRODUCTION

The understanding of the underlying mechanisms of cancer is one of the great challenges of modern medicine. According to the latest estimates on the global burden of cancer, released by the International Agency for Research on Cancer (IARC), 18.1 million new cases and 9.6 million deaths have emerged in 2018 [1]. Despite the efforts that are being made to understand the many processes and mechanisms driving cancer and also to improve the efficacy of current treatments, much remains to be done to fight this lethal disease. Mathematical modeling is a powerful tool that is extensively used to describe real-world problems illuminating different disciplines of science and engineering, including oncology [2]. Several oncological mathematical models have proven to be very useful in providing guiding principles for better illuminating the complex biological processes and mechanisms underlying cancer growth. In addition, they have shown to be of potential help in the quest of optimal therapy scheduling [3], for the ultimate purpose of approaching towards personalized cancer therapy.

[1]Zehor Belkhatir, James C. Mathews, Maryam Pouryahya and Joseph O. Deasy are with Department of Medical Physics, Memorial Sloan Kettering Cancer Center, NY {belkhatz,mathewj2,pouryam1,deasyj}@mskcc.org

[2]Michele Pavon is with Department of Mathematics, University of Padova, Italy pavon@math.unipd.it

[3]Larry Norton is with Department of Medicine, Memorial Sloan Kettering Cancer Center, NY nortonl@mskcc.org

[4]Allen R. Tannenbaum is with Departments of Computer Science and Applied Mathematics & Statistics, Stony Brook University, NY arobertan@cs.stonybrook.edu

Different mathematical models for tumor growth have been proposed in the last decades. Those models can be arranged in a broad spectrum ranging from simple macroscopic models (with fewer details and fewer parameters) trying to emulate the clinically observed growth of tumor volume to more sophisticated models involving the microscopic and/or molecular processes that contribute to tumor growth [4], [5], [2]. A plethora of macroscopic tumor growth models, based on ordinary differential equations (ODEs) exist in the literature including those based on logistic, Gompertezian, and Von Bertalanffy models; see [6], [2] and the references therein. However, it should be stressed that often real tumor growth data is not smooth and discrepancies exist between the measured data and the ones simulated from theoretical deterministic models. In fact, deterministic models do not account for the different sources of variability, often not quantifiable or even unknown, that are natural to the process of tumor growth. As examples of such sources of variability, one may think about the intrinsic heterogeneity of tumoral tissues and the environmental stochasticity resulting from random fluctuations in the environment. A possible approach to take into account the different sources of variability in tumor growth models is to add a noise term to a given deterministic ODE model, which leads to a stochastic differential equation (SDE) model. This class of models has been used to describe the tumor growth process, e.g., in [7], [8].

Modeling the effect of cytotoxic chemotherapy on tumor growth dynamics has seen several important developments in the past several decades. Furthermore, the question of identifying the optimal anti-cancer chemotherapeutic treatment for a specific patient is fundamental, yet challenging to answer conclusively. The current clinical practice for scheduling chemotherapy follows certain accepted standards regarding the choice of the prescribed drug's dose and administration time, usually adjusted according to the phase of the tumor, patient's body weight, white blood cell level, among other factors. The limitation of this largely empirical approach has increasingly been recognized in the scientific and clinical communities. There is, therefore, a need to optimize current chemotherapy planning for cancer treatment to limit toxicities, advance clinical trial pathways for new drugs, and enable personalized therapy. Towards this end, approaches based on control systems engineering have been developed for the automation of chemotherapy treatment. These *in silico* trials do not need the huge costs of laboratory experiments and can also be tested in a relatively fast time. A review of the mathematical-based techniques applied to the optimal cancer therapy planning is provided in [9], [10]. Other feedback control strategies were investigated, e.g., in [11], [12], [13] and references therein.

Even though several techniques have been proposed to address the control and identification problems associated with different tumor growth models, to the best of our knowledge, this is the first study that considers the analysis of a stochastic version of the so-called "Norton-Simon-Massagué" (NSM) tumor growth model, a biologically grounded minimal order model, for the purpose of estimation and control. We aim ultimately to find general guiding principles from the proposed stochastic NSM tumor growth model using tools from identification and control theories. In this regard, we (i) propose a stochastic extension of the deterministic NSM tumor growth model; (ii) derive conditions under which the tumor volume remains positive; (iii) calibrate the proposed stochastic model using a population mixed-effect formulation and an estimation algorithm that relies on maximum likelihood estimator (MLE) and Extended Kalman Filter (EKF) techniques; and (iv) study and discuss the effect of adding the chemotherapeutic control agent into the unperturbed stochastic model both in open-loop and closed-loop stand-points. The closed-loop control strategy that is used to solve an optimal cancer therapy scheduling problem relies on the combination of model predictive control (MPC) and EKF methods.

The remainder of this paper is organized as follows. Section III describes a general stochastic extension of the deterministic NSM tumor growth model in the absence of therapy, along with the positivity proof of its solution. In Section IV, the estimation algorithm used to calibrate the proposed unperturbed stochastic model is provided along with some simulation results showing the performance of the algorithm. The effect of the chemotherapy drug agent is included in the unperturbed model in Section V where an analysis of the controlled model is performed in open-loop and closed-loop as well. Concluding remarks and future research directions are provided in Section VI.

## II. Notations

Throughout the paper, unless otherwise specified, the following notation will be used. The probability space is denoted by $(\Omega, \mathscr{F}, \mathbb{P})$, where $\Omega$ is the set of all possible outcomes, $\mathscr{F}$ is a $\sigma$-algebra of events with a filtration $\{\mathscr{F}_t\}_{t \geq 0}$, and $\mathbb{P}$ is the probability measure function of the event. Let $W(t)$ be the one-dimensional Brownian motion defined on the probability space. The space $\mathbb{R}$ and $\mathbb{R}_+^* = (0, +\infty)$ denotes the usual spaces of real and positive numbers, respectively; $C^2(X; S)$ is the space of all continuous functions $f : X \to S$ with continuous second derivatives. The notation a.s. stands for almost surely. The operator $\mathbb{E}$ denotes the expected value, $\tau_n \wedge T = \min(\tau_n, T)$, and mod $(.,.)$ denotes the modulo operation. The characteristic function on $X$ is denoted $\chi(X)$.

## III. Uncontrolled stochastic norton-simon-massagué tumor growth model

In this section, we present a stochastic extension of the uncontrolled deterministic NSM model, also known as Von Bertalanffy growth model, to account for the different sources of random effects that are inherent in the complex and heterogeneous process of tumor growth. We also provide conditions on the model's parameters that preserve positivity of the tumor volume.

### A. Model description

The Von Bertalanffy model was proposed in the fifties to describe the growth of biological organisms based on basic energetics principles [14]. The derived growth model is given as follows,

$$\frac{dV(t)}{dt} = aV^\alpha(t) - bV(t), \tag{1}$$

where $a$ and $b$ are constants of anabolism (growth) and catabolism (death) respectively. Equation (1) states that the net growth rate of an organism results from the balance of synthetic and degradative processes. While the rate of the former process follows a law of allometry (i.e., the rate is proportional to the volume $V(t)$ via a power function), the rate of the latter process scales linearly with $V(t)$.

The two special cases of (1), (i) power law $b = 0$, and (ii) second type growth $\alpha = 2/3$ (or often termed as Von Bertalanffy as well) have already been successfully applied to describe tumor growth [15], [16]. The general case, $0 < \alpha < 1$, was introduced in [17] to explain the

 

self-seeding hypothesis. Moreover, a geometrical interpretation was provided in [18], [6], which relates the exponent $\alpha = d/3$ to the fractional Hausdorff dimension of the proliferative tissue, where $d$ denotes the fractal dimension of proliferative tissue. Furthermore, in 2011, model (1) was derived mechanistically by linking tumor growth to metabolic rate and vascularization [19]. In the rest of the paper, we will refer to model (1) as the deterministic NSM tumor growth model[1].

To consider the intertumor and intratumor heterogeneity in the tumor growth process, we propose to extend the deterministic ODE model (1) to a more general stochastic version, where the different types of randomness are modeled in the diffusion term of (2). The time evolution of the tumor volume in the absence of therapy is described by the following SDE of Itô-type:

$$\begin{cases} dV(t) = (aV^{\alpha}(t) - bV(t)) \, dt + \sigma V^{\beta}(t) dW(t), \\ V(0) = V_0, \end{cases} \tag{2}$$

driven by a standard Wiener process $(W(t) : t \geq 0)$ and started at $V_0 \in \mathbb{R}_+^*$, where $a, b, \alpha, \sigma, \beta$ are positive real parameters. The proposed stochastic generalization (2) encompasses the cases of $\beta = \alpha$ and $\beta = 1$ that can be interpreted as random versions resulting from randomly perturbing the growth and death parameters ($a$ and $b$) of the deterministic model by Gaussian distributions.

### B. Regularity of the solution on $\mathbb{R}_+^*$

An important property that makes SDE (2) valid in describing the dynamics of tumor volumes is the invariance of the solution $(V(t) : t \geq 0)$ with respect to $\mathbb{R}_+^*$ (set of all strictly positive real numbers). Theorem 1 provides conditions on the parameters of model (2) that guarantee the positivity of the solution $V(t)$.

**Theorem 1.** *Let $V_0 \in \mathbb{R}_+^*$ be any given initial condition. If the conditions $0 < \alpha < 1$, $b \geq 0$; and $\left\{\beta > \frac{\alpha+1}{2}, a \geq 0, \sigma \geq 0\right\}$ or $\left\{\beta = \frac{\alpha+1}{2}, \sigma^2 \leq 2a\right\}$ hold, then the solution $V(t)$, for $t \geq 0$, of the SDE (2) is regular on $\mathbb{R}_+^*$, i.e., $\mathbb{P}\left(V(t) \in \mathbb{R}_+^*\right) = 1$ for all $t \geq 0$.*

*Proof:* First, let us denote the drift and diffusion terms of the SDE (2) as follows:

$$k(x) = ax^{\alpha} - bx, \qquad \lambda(x) = \sigma x^{\beta}, \tag{3}$$

---

[1] The NSM name was given to refer to the self-seeding concept introduced by L. Norton and J. Massagué, and also to indicate the "Norton-Simon" hypothesis, which was proposed by L. Norton and R. Simon, and which will be used to model the chemotherapeutic drug agent later on in this study

Given that the drift and diffusion terms defined in (3) are locally Lipschitz on $\mathbb{R}_+^*$, there exists a unique local solution $V(t)$ on $t \in [0, \tau_e)$, where $\tau_e$ is the explosion time [20]. In order to prove that the solution is global, we need to show that $\tau_e = \infty$ a.s. In the following analysis, we will distinguish among three different cases. In the first case, we make use of Lyapunov-type methods, and in the second and third cases we invoke the classical Feller's test for explosion [20]. These techniques have been used to prove positivity results of different stochastic models, see, e.g., [20].

- **Case 1:** $\beta \geq 1$

  Let $n_0$ be a sufficiently large integer, such that any given $V_0 \in \mathbb{R}_+^*$ lies within the interval $\mathbb{D}_{n_0} = \left( \frac{1}{n_0}, n_0 \right)$. Given any integer $n \geq n_0$, we define the stopping time as follows:

  $$\tau_n = \inf \{ t \in [0, \tau_e) : V(t) \notin \mathbb{D}_n \},\tag{4}$$

  with $\inf\{\emptyset\} = \infty$. The sequence $\{\tau_n\}$ is increasing when $n \to \infty$ with a limit denoted as $\tau_\infty = \lim\limits_{n \to \infty} \tau_n$. It is clear that $\tau_\infty \leq \tau_e$ a.s. Thus, if $\tau_\infty = \infty$ a.s., then $\tau_e = \infty$ a.s., and $V(t) \in \mathbb{R}_+^*$ a.s., for $t \geq 0$. As a consequence, we need to prove that $\tau_\infty = \infty$ a.s. We will proceed by contradiction. If $\tau_\infty \neq \infty$ a.s, then there exist two constants $T > 0$ and $\varepsilon \in (0, 1)$ such that

  $$\mathbb{P}(\tau_\infty \leq T) > \varepsilon,\tag{5}$$

  and as a consequence there exists an integer $n_1 \geq n_0$ such that

  $$\mathbb{P}(\tau_n \leq T) \geq \varepsilon, \qquad n \geq n_1.\tag{6}$$

  Next we define $F \in C^2(\mathbb{R}_+^*; [0, \infty))$ as follows:

  $$F(x) = \sqrt{x} - 1 - 0.5\ln(x).\tag{7}$$

  Applying the Itô formula on the function $F(V(t))$ yields, for $V(t) \in \mathbb{R}_+^*$

  $$dF(V(t)) = \mathscr{L}_1 F(V(t))dt + \mathscr{L}_2 F(V(t))dW(t),\tag{8}$$

  where

  $$\begin{cases} \mathscr{L}_1 F(V(t)) = 0.5a\left(V^{\alpha-0.5} - V^{\alpha-1}\right) + 0.5b\left(1 - V^{0.5}\right) + 0.25\sigma^2\left(V^{2\beta-2} - 0.5V^{2\beta-1.5}\right), \\ \mathscr{L}_2 F(V(t)) = 0.5\sigma\left(V^{\beta-0.5} - V^{\beta-1}\right). \end{cases}\tag{9}$$

Since the coefficients of the highest and lowest order of $\mathscr{L}_1 F(V(t))$ are both negative (they are $-0.125\,\sigma^2$ and $-0.5a$ respectively), then $\mathscr{L}_1 F(V(t))$ is bounded by a constant $c_1$, on $V \in \mathbb{R}_+^*$. Therefore, we get

$$dF(V) \leq c_1\,dt + \mathscr{L}_2 F(V(t))\,dW(t). \tag{10}$$

Integrating both sides of (10) from 0 to $\tau_n \wedge T$, and then taking the expected value of the resulted equation yields,

$$\mathbb{E}\left[F(V(\tau_n \wedge T))\right] \leq F(V_0) + c_1\,\mathbb{E}\left[\tau_n \wedge T\right] \leq F(V_0) + c_1\,T, \tag{11}$$

where

$$
\begin{aligned}
\mathbb{E}\left[F(V(\tau_n \wedge T))\right] &= \mathbb{E}\left[\chi(\tau_n \leq T)\,F(V(\tau_n \wedge T))\right] + \mathbb{E}\left[\chi(\tau_n > T)\,F(V(\tau_n \wedge T))\right] \\
&\geq \mathbb{E}\left[\chi(\tau_n \leq T)\,F(V(\tau_n))\right],
\end{aligned}
\tag{12}
$$

with $\chi(.)$ is the characteristic function.

Define $\Omega_n = \{\tau_n \leq T\}$, for $n \geq n_1$. Then by virtue of (6), $\mathbb{P}(\Omega_n) \geq p$. So, for $\tau_n \in \Omega_n$, $V(\tau_n) = n$ or $\frac{1}{n}$, and thus,

$$F(V(\tau_n)) \geq \left[F(n) \wedge F\left(\frac{1}{n}\right)\right]. \tag{13}$$

Therefore, from (11)-(13) we get

$$F(V_0) + c_1\,T \geq \mathbb{E}\left[F(V(\tau_n \wedge T))\right] \geq \mathbb{E}\left[\chi(\Omega_n)\,F(V(\tau_n))\right] \geq \tau\left[F(n) \wedge F\left(\frac{1}{n}\right)\right]. \tag{14}$$

Letting $n \to \infty$, results in

$$\infty > F(V_0) + c_1\,T \geq \infty, \tag{15}$$

which is a contradiction. So, the starting hypothesis is false, i.e., $\tau_\infty = \infty$ a.s., and thus $\tau_e = \infty$ a.s.

- **Case 2:** $\frac{\alpha+1}{2} < \beta < 1$

Define the following quantities:

$$g(x) \triangleq \int_1^x \exp\left\{-2\int_1^y \frac{k(z)}{\lambda^2(z)}\,dz\right\}dy, \tag{16}$$

where $k(z)$ and $\lambda(z)$ are the drift and diffusion coefficients defined in (3).

After some computations, we get

$$g(x) = e^{m_1}\int_1^x \exp\left\{\frac{-2a}{\sigma^2(\alpha - 2\beta + 1)}y^{\alpha - 2\beta + 1} + \frac{b}{\sigma^2(1-\beta)}y^{2(1-\beta)}\right\}dy, \tag{17}$$

where $m_1 = \dfrac{b}{\sigma^2(1-\beta)} - \dfrac{2a}{\sigma^2(\alpha-2\beta+1)}$.

Using the quotient test for integrals with non-negative integrands, along with the conditions $\alpha - 2\beta + 1 < 0$, $\beta < 1$, $a \geq 0$ and $b \geq 0$, we can show that $\lim_{x \to 0^+} g(x)$ and $\lim_{x \to \infty} g(x)$ are divergent integrals. Moreover, given that the function $g(x)$ is increasing on $\mathbb{R}_+^*$, we conclude that:

$$\lim_{x \to 0^+} g(x) = -\infty, \qquad \lim_{x \to \infty} g(x) = \infty. \tag{18}$$

Thus, Feller's test conditions hold. Hence, $\mathbb{P}(\tau_\infty = \infty) = 1$.

- **Case 3:** $\beta = \frac{\alpha+1}{2}$

  Following the same computation steps as in case 2, we get

  $$g(x) = e^{m_2} \int_1^x y^{-\frac{2a}{\sigma^2}} \exp\left\{\frac{2b}{\sigma^2(1-\alpha)} y^{1-\alpha}\right\} dy \tag{19}$$

  where $m_2 = \dfrac{b}{\sigma^2(1-\beta)} - \dfrac{2a}{\sigma^2(\alpha-2\beta+1)}$.

  The Feller's test conditions (18) are satisfied if $2a \geq \sigma^2$. Hence, $\mathbb{P}(\tau_\infty = \infty) = 1$.

  ∎

**Remark 1.** *The cases of $\alpha = 1$ and $\alpha = 0$ correspond to the theta process and mean reverting theta process respectively. These models have been used to describe, e.g., many applications from economics and finance. The positivity result of their solutions has been proved using similar techniques and can be found in Chapter 9 of [20].*

## IV. Calibration of the uncontrolled stochastic NSM growth model

The stochastic NSM tumor growth model is usually not entirely known *a priori* since information about the model's parameters may be missing. It is, therefore, essential to develop an efficient estimation algorithm to extract estimates of the growth and noise parameters from the available tumor volume measurements. In this section, details about a maximum likelihood estimator (MLE) based algorithm are provided along with numerical results showing its performance.

## A. Population mixed-effect modeling approach

When only sparse data measurements are available for several individuals from a population, which is the case for tumor volume data, one can use the population mixed-effect modeling approach [21] to estimate the parameters of the investigated model. This framework encompasses two stages. The first stage describes the intra-individual variability, while the second stage models the inter-individual variations. The calibration of such a class of models is performed by pooling the time series data of all subjects into a single larger measurement set which may potentially resolve identifiability issues when only a few data points are available per subject.

Let us consider the following population mixed-effect NSM tumor Growth model:

$$\begin{cases} dV_i(t) = \left(aV_i^{\alpha}(t) - bV_i(t)\right)dt + \sigma V_i^{\beta}(t)\,dW(t), & i = 1,2,...,N, \\ y_{ij} = V_i(t_j) + \varepsilon_{ij}, & j = 1,2,...,M, \\ V_i(0) = V_{i0}, \end{cases} \tag{20}$$

where $i = 1,2,...,N$ denotes the number of individuals in a population, $j = 1,2,...,M$ represents the number of time point measurements and $\varepsilon_{ij} \sim \mathcal{N}(0,S)$ is the measurement noise assumed to be Gaussian with zero mean and covariance $S$. In this work, we assume that the inter-individual variability comes from the initial condition which is given as follows

$$G(V_{i0}) = G(V_{io}^M) + \eta_i, \qquad \eta_i \sim \mathcal{N}(0,\Omega_1). \tag{21}$$

where $G(.)$ is a nonlinear mapping defined in Remark 2, $V_{i0}^M$ is the fixed effect initial condition for the volume, and $\eta_i$ is the random effect parameter that describes the variation of each individual from the population value $V_{i0}^M$. Thus, the full model's parameters to be estimated are defined in $\Theta = (\theta, \Omega_1)$, where $\theta = (a, b, \alpha, \sigma, S, \beta, V_{i0}^M)$ is the vector of the fixed effect parameters.

## B. Estimation algorithm

The estimation algorithm used in this paper is based on the MLE combined with the EKF. We provide in this subsection the main steps of the estimation algorithm, and for more details we refer the reader to [22] and references therein.

The population likelihood function for fixed-effects is defined as

$$L(\theta) = \prod_{i=1}^{N} \int p(\mathscr{Y}_i | \theta, \eta_i) p(\eta_i | \Omega_1)\,d\eta_i = \prod_{i=1}^{N} \int \exp(l_i)\,d\eta_i \tag{22}$$

where $N$ is the number of individuals, $\mathscr{Y}_i$ is the vector of all observations for the $i^{th}$ individual, $(\theta, \eta_i, \Omega_1)$ are the parameters defined in the previous subsection, $p(\mathscr{Y}_i | \theta, \eta_i)$ is the probability for the first stage model, $p(\eta_i | \Omega_1)$ is the probability related to the second stage that relates the random effects to the inter-individual variation, and $l_i$ is a posteriori log-likelihood function.

Often the population likelihood function (22) cannot be computed analytically and therefore needs to be approximated. In [22], the *a posteriori* log-likelihood function $l_i$ is approximated by a second-order Taylor expansion around $\eta_i^*$, the maximizer of $l_i$. The population likelihood function to be maximized is given as follows [22]:

$$L(\theta) \approx \prod_{i=1}^{N} \left| \frac{-\Delta l_i^*}{2\pi} \right|^{-(1/2)} \exp(l_i)|_{\eta_i^*}, \tag{23}$$

where the second derivative $\Delta l_i^*$ is approximated using first-order conditional estimation method:

$$\Delta l_i^* \approx -\sum_{j=1}^{n_i} (\nabla \varepsilon_{ij}^T R_{i(j|j-1)}^{-1} \nabla \varepsilon_{ij}) - \Omega_1^{-1}, \tag{24}$$

with $\nabla \varepsilon_{ij} = \frac{\partial}{\partial \eta_i} \varepsilon_{ij}|_{\eta_i^*}$, and the output error residual $\varepsilon_{ij}$ and conditional residual covariance $R_{i(j|j-1)}$ of individual $i$ are computed using the EKF.

The optimization of the population likelihood function (23) is nested in the sense that approximation (24) depends on the optimal random effect $\eta_i^*$, the optimizer of $l_i$. This estimation approach is implemented in the statistical programming language R within Population Stochastic Modeling (PSM) package [22].

**Remark 2.** *Because the EKF cannot handle SDEs with state-dependent diffusion terms, we apply Lamperti transformation [23] to remove the state dependence from the diffusion term of* (20). *The resulting transformed model is given as follows, for $i = 1, 2, ..., N$, and $j = 1, 2, ..., M$,*

- *Case 1: $\beta \neq 1$:*

$$\begin{cases} dZ_i(t) = (1-\beta) Z_i(t) \left( a Z_i^{\frac{\alpha-1}{1-\beta}}(t) - b \right) + \frac{\sigma^2}{2} \beta(\beta-1) Z_i^{-1}(t) \, dt + \sigma(1-\beta) \, dW(t), \\ y_{ij} = Z_i^{\frac{1}{1-\beta}}(t_j) + \varepsilon_{ij}, \end{cases} \tag{25}$$

  *where $Z_i(t) = V_i^{1-\beta}(t)$.*

- *Case 2: $\beta = 1$:*

$$\begin{cases} d\bar{Z}(t) = \left( a e^{(\alpha-1)\bar{Z}_i} - b - \frac{\sigma^2}{2} \right) dt + \sigma \, dW(t), \\ y_{ij} = e^{\bar{Z}(t_j)} + \varepsilon_{ij}, \end{cases} \tag{26}$$

                                                          

where $\bar{Z}_i(t) = \ln[V_i(t)]$, and $\ln[.]$ is the natural logarithmic function .

## C. Results

*1) Data description:* We consider tumor volume data from two different in vivo experiments: an orthotopically xenografted human breast carcinoma and an ectopic syngeneic tumor (Lewis lung carcinoma). These data were published in [6], and for more details about mice experiments, one can refer to the latter reference. The lung data were downloaded from the webpage of Sébastien Benzekry[2], however, the breast data were extracted from figures published in the supplementary document of [6] using WebPlotDigitizer[3]. The temporal profiles of the breast and lung data are depicted in Figure 1, where both groups contain eight subjects.

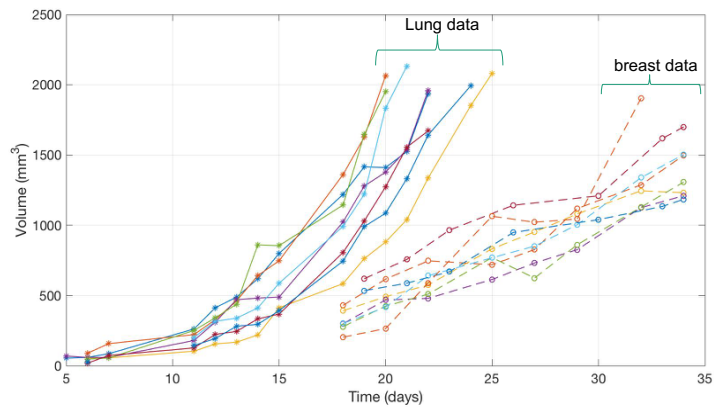

Fig. 1: Populations of lung and breast mice data.

*2) Fitting results:* The calibration of the growth model (26) with multiplicative noise is performed by fitting the data shown in Figure 1. The initial values of the algorithm were selected through a trial and error strategy, and the available volume data is specified to the algorithm in the log-domain.

The values of the parameter estimates and their 95 % confidence intervals (CIs) are given in Tables I and II for the lung and breast datasets respectively. Compared to the CIs of the model's

---

parameters using lung data, the ones corresponding to the breast data are wider. Interestingly though, the estimate of the noise amplitude $\sigma$ with its CI is comparable for both data populations.

| Parameters / Estimates | $a$ | $b$ | $\alpha$ | $V_{i0}^M$ | $\sigma$ | $S$ | $\Omega_1$ |
|---|---|---|---|---|---|---|---|
| Lower bound 95% CI | 0.4765 | 0.0199 | 0.8028 | 26.5146 | 0.0439 | 0.01107 | 0.0456 |
| MLE | 0.6851 | 0.1185 | 0.8857 | 36.1776 | 0.08511 | 0.02225 | 0.14995 |
| Upper bound 95% CI | 0.98 | 0.7054 | 0.9685 | 49.3623 | 0.1263 | 0.0334 | 0.4923 |

TABLE I: Parameter estimates with the lower and upper bounds of the 95% confidence interval using lung data measurements.

| Parameters / Estimates | $a$ | $b$ | $\alpha$ | $V_{i0}^M$ | $\sigma$ | $S$ | $\Omega_1$ |
|---|---|---|---|---|---|---|---|
| Lower bound 95% CI | 0.7365 | 0.0219 | 0.3171 | 279.5808 | 0.037 | $-0.006$ | 0.0243 |
| MLE | 1.9638 | 0.3671 | 0.7822 | 344.1297 | 0.0864 | 0.0038 | 0.0768 |
| Upper bound 95% CI | 5.2361 | 6.1352 | 1.2474 | 423.5815 | 0.1358 | 0.0137 | 0.2426 |

TABLE II: Parameter estimates and their 95% confidence interval using breast data measurements.

The behavior of the two identified models using their corresponding estimated parameter values was examined through the output one-step predicted and smoothed estimates as shown in Figures 2 and 3. Both types of state estimates are based on the EKF, where the predicted estimate uses observation up to $t_{k-1}$ to predict the output at $t_k$ and the smoothed estimate uses all observations at each time point. The smoothed output is the optimal type of estimates because all the available data is used to estimate the tumor volume at a specific time instant, and this can be clearly observed from Figures 2 and 3. The one-step prediction is less accurate in the breast data compared to the lung data.

Overall, the calibration of the population stochastic NSM tumor growth model employing breast data is less accurate compared to the identified model using lung data. One main reason may be due to the additional bias introduced when collecting the data manually from published figures, and another reason may be related to the fewer data points that are available for each individual in that particular population.

## V. Controlled stochastic Norton-Simon growth model

The log-kill hypothesis, which posits that a given dose of chemotherapy kills the same fraction of tumor cells regardless of the size of the tumor at the time of administration [24], has for decades guided the clinical treatment of many types of cancer. This model led to the administration of maximum tolerated dose (MTD) of a cytotoxic agent with prolonged treatment breaks to counteract disease progression and to kill as many cancer cells as possible while allowing the body to recover from the induced treatment toxicity. This regimen worked experimentally for haematological cancers (leukemia) [24], but it failed with many types of solid tumors. Based on clinical observations that were not in agreement with the outcomes of log-kill hypothesis, the Norton-Simon hypothesis emerged [25], [26]. It states that cancer cell death in response to a chemotherapeutic drug agent is proportional to the untreated tumor growth rate at the time of treatment. This model led to the finding that not only dose intensity is important but also dose density. Different clinical trials have been conducted to validate this hypothesis [27].

In this section, we model the effect of chemotherapy drug agent into the unperturbed stochastic growth model (2) following the Norton-Simon hypothesis, and we analyze the perturbed model both in open and closed loops.

### A. Mathematical model

Mathematically, the controlled stochastic NSM model is formulated by the following equation:

$$dV(t) = (aV^{\alpha}(t) - bV(t))(1 - \xi C(t)) \, dt + \sigma V^{\beta}(t) \, dW(t), \tag{27}$$

where $\xi$ reflects the sensitivity of the tumor to the drug whose concentration is denoted $C(t)$. The dynamics of the drug concentration $C(t)$ is described using the one-compartment pharmacokinetics model [28] given as follows

$$\frac{dC(t)}{dt} = -kC(t) + u(t), \tag{28}$$

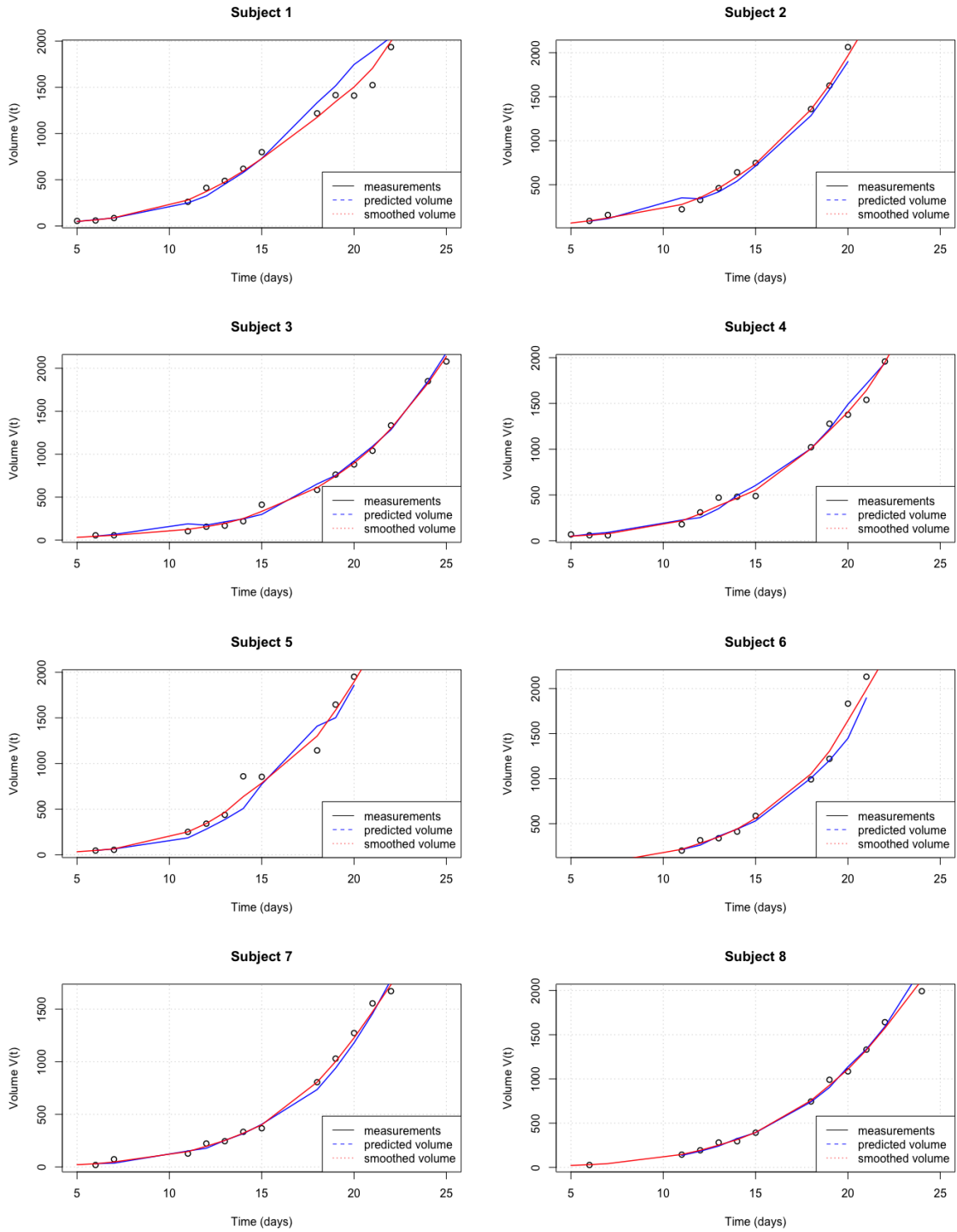

Fig. 2: Fitted tumor volumes for subjects within the lung tumor population. Black circles: real tumor volume measurements; blue lines: one-step predicted tumor volumes; and red lines: smoothed tumor volumes.

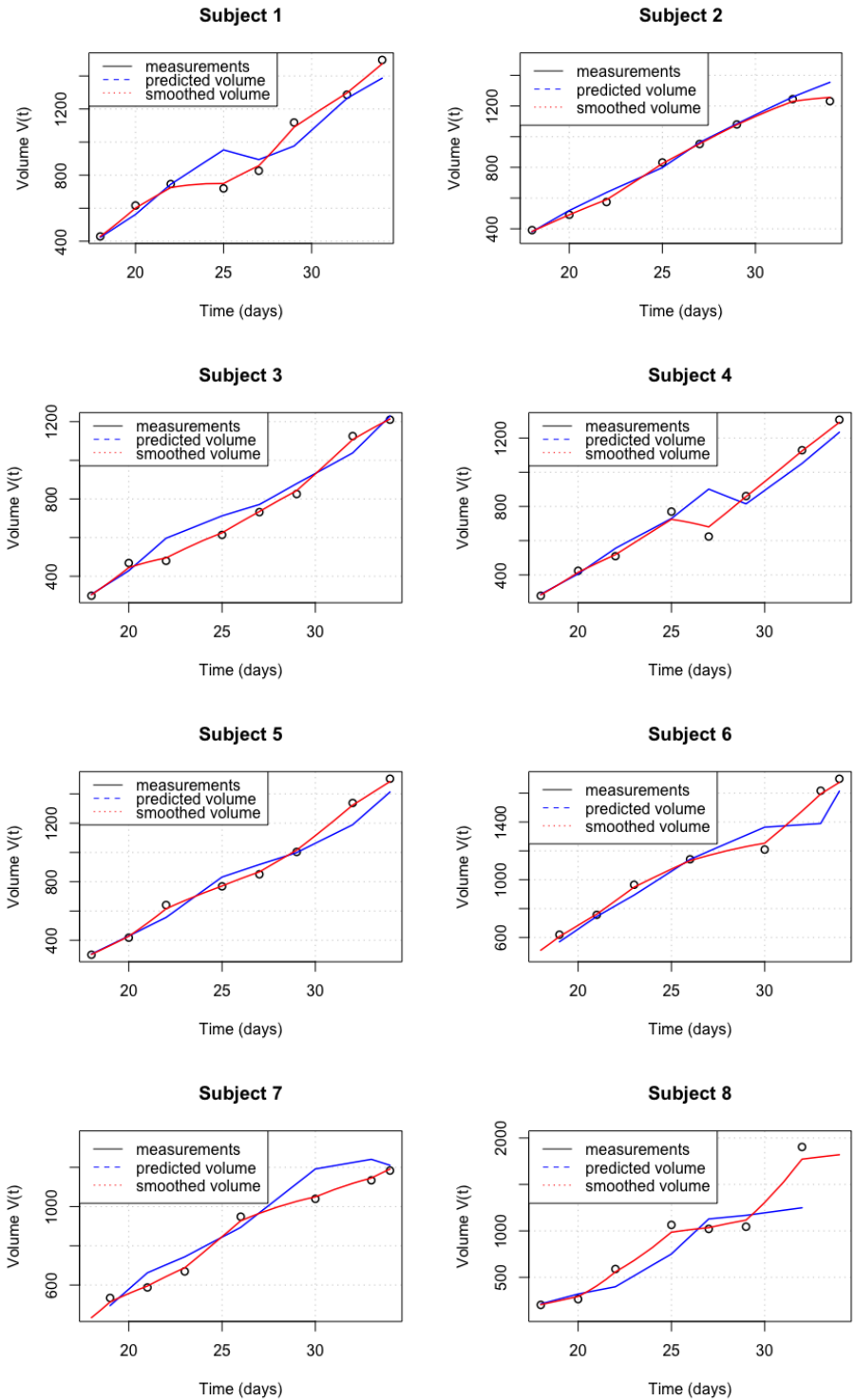

Fig. 3: Fitted tumor volumes for subjects within the lung tumor population. Black circles: real tumor volume measurements; blue lines: one-step predicted tumor volumes; and red lines: smoothed tumor volumes.

where $u(t)$ represents the rate of flow of drug into the body, and $k$ is the elimination rate constant of the drug from bloodstream. The elimination half-life of the drug can be computed as $t_{1/2} = \ln(2)/k$.

### B. Analysis in open-loop

We will now conduct a comparative study between different protocols of chemotherapy. The analysis is performed in open-loop where the drug profiles are fixed *a priori* and does not change with the measured volume. The mathematical description of the drug infusion is given as follows [29]

$$u(t) = \begin{cases} \dfrac{D}{t_i} & \text{if } \mathrm{mod}(t,T) \in [0, t_i), \\ 0 & \text{if } \mathrm{mod}(t,T) \in [t_i, T), \end{cases} \qquad (29)$$

where $D$ is the dose of the drug in the course of chemotherapy, $t_i$ is the time during which the drug is given, and $T$ is the time between cycles.

Four different chemotherapeutic protocols are compared. A schematic diagram of a typical protocol of chemotherapy is depicted in Figure 4. The two main characteristics being compared in those chosen protocols are dose-intensity and dose-density while assuming the elimination rate of the cytotoxic drug from bloodstream $k$ and the duration of infusion $t_i$ are constant. The first protocol, which is used, e.g., for breast cancers[4], is considered as our reference with parameter values $T = 3$ weeks, $t_i = 1$ hour, and $D = 60 \, \mathrm{mg}\,\mathrm{m}^{-2}$. The second and third protocols are more dose-dense compared to the first one, with $T = 2$ weeks and $T = 1$ week, respectively, and the dose-intensity $D$ is kept the same. The last protocol tests the effect of densifying even more the therapy plan while lowering the intensity, where $T = 2$ days and $D = 20 \, \mathrm{mg}\,\mathrm{m}^{-2}$. We emphasize that the total dose per course treatment is constant in the four protocols. The temporal profiles of the injected drug and its concentration at the tumor site are depicted in Figure 5 for the four therapy plans. We refer to the four plans as Maximum Tolerated Dose (MTD), Dense MTD (DMTD), High DMTD (HMTD), and Low dose metronomic (LDM) respectively.

---

[4]Standard protocols of chemotherapy can be found in http://www.bccancer.bc.ca/health-professionals/clinical-resources/chemotherapy-protocols

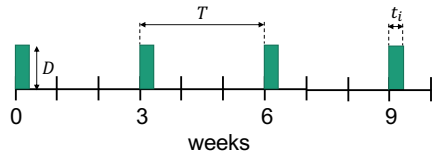

Fig. 4: A typical chemotherapy protocol: $D$ dose intensity, $T$ time between cycles, $t_i$ infusion time.

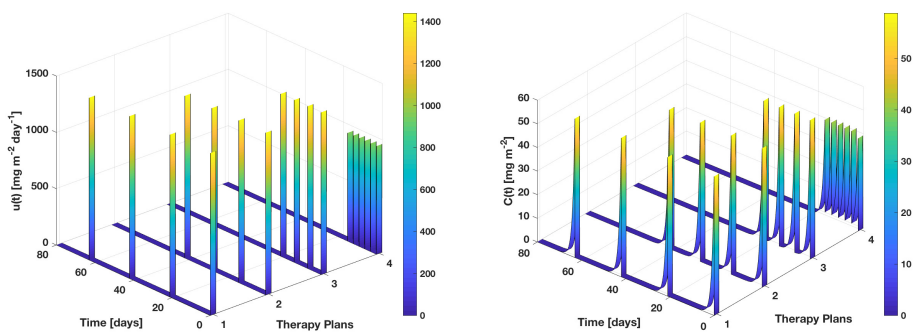

Fig. 5: Infused dose $u(t)$ and its concentration at tumor site $C(t)$.

The objective of this subsection is to investigate the sensitivity of the response to different chemotherapeutic protocols to the heterogeneity in tumor growth dynamics, namely the parameters $\sigma\,[\mathrm{mm}^{3(1-\beta)}\,\mathrm{day}^{-1}]$ and $\beta\,[-]$, for two scenarios of tumor growth. The high rate of growth scenario (higher fractal dimension) corresponds to a value $\alpha = 2.98/3\,(d = 2.98)$, while the low rate of growth scenario (lower fractal dimension) corresponds to a value $\alpha = 2.25/3\,(d = 2.25)$[5]. All chemotherapy plans start at $t = 0$, where the initial volume is $V(0) = 1000\,\mathrm{mm}^3 \sim 10^9$ cells corresponding to the size assumed to be clinically detectable. The values of the fixed parameters in the four protocols are: $a = 0.6851$ $\mathrm{mm}^{3(1-\alpha)}\,\mathrm{day}^{-1}$, $b = 0.1185$ $\mathrm{day}^{-1}$, $k = 1\,\mathrm{day}^{-1}$, and $\xi = 0.3$ $\mathrm{mg}^{-1}\,\mathrm{m}^2\,\mathrm{day}$. The growth and death parameters are taken to be the estimates derived in Section IV using the lung data.

---

[5]These values appear in [18] and correspond to the fractal dimension in infiltrating ductal adenocarcinoma and in normal breast tissue, respectively

The results we obtained for the two investigated growth type scenarios are shown in Figures 6 and 7. We organize the results in three layers, where each layer answers a particular question resulting from the answer to the question of the previous layer. The first layer starts with a stochastic growth model perturbed with a multiplicative noise ($\beta = 1$) having a certain amplitude ($\sigma = 0.02$), which is the same model identified in Section IV. The first question we pose is: *"Q1: What is the response, for that specific model, of the different chosen therapy plans?"*. Answering the first question allows us to move to the second layer of figures where the question of interest is: *"Q2: How does the response to different therapies change with different noise levels, but still multiplicative noise?"*. The answer to *Q2* allows us to find the best value of noise level where the therapies are more effective, and leads to the last question investigated in the last layer: *"Q3: Can the response to the different treatments, for the best value of noise amplitude, be even more effective if the stochasticity follows a low of allometry instead of being multiplicative?"*. This set of questions study the sensitivity of stochastic parameters on the controlled tumor growth model. Answers (*A1-A3*) to these questions are provided within the captions of Figures 6 and 7.

From the extensive numerical investigations that we conducted to study the sensitivity of therapy response to the stochastic parameters $\sigma$ and $\beta$, the observation which is worth being emphasized is that dose-dense protocols are generally more beneficial.

### C. Closed-loop optimal dosing strategy

In this subsection, we aim to solve an OCP in a feedback manner to plan chemotherapy for cancer treatment. In order to do so, we use the MPC approach combined with the EKF for state estimation. The MPC method, also known as *receding horizon control*, is a feedback control strategy that was developed since the seventies and has been successfully applied to various fields [30]. The control input in MPC is determined by solving at each sampling instance an open-loop finite-horizon OCP.

To design the control therapy plan, we use the same uncontrolled model identified in Section IV combined with a cytotoxic drug effect. The state-space representation of the controlled stochastic NSM tumor growth model to be used in the OCP formulation is given as follows:

$$\begin{cases} dx(t) = f(x,u)\,dt + g(\sigma)\,dW(t), \\ \bar{y}_j = \ln[y_j] = x_1(t_j) + \varepsilon_j, \end{cases} \tag{30}$$

 

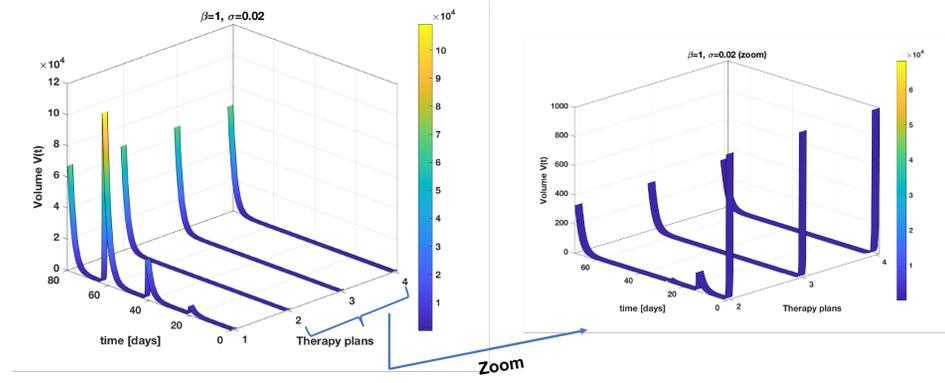

(a)

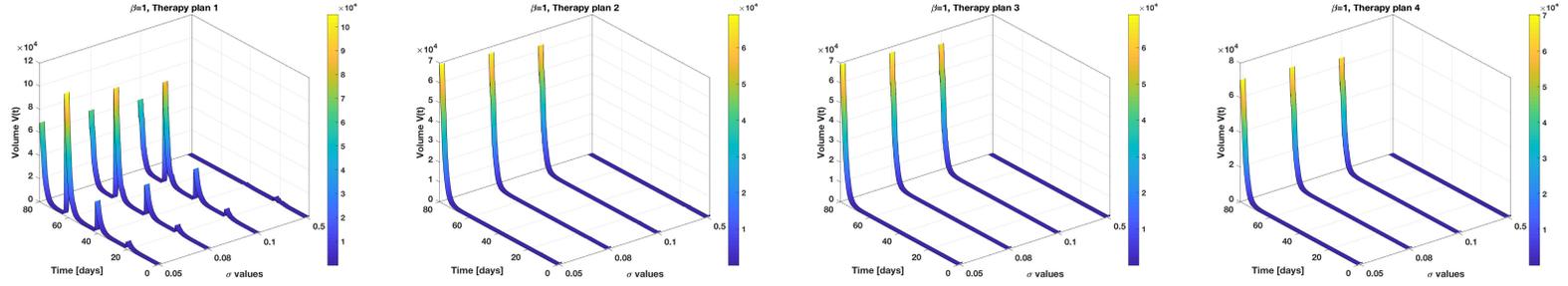

(b)

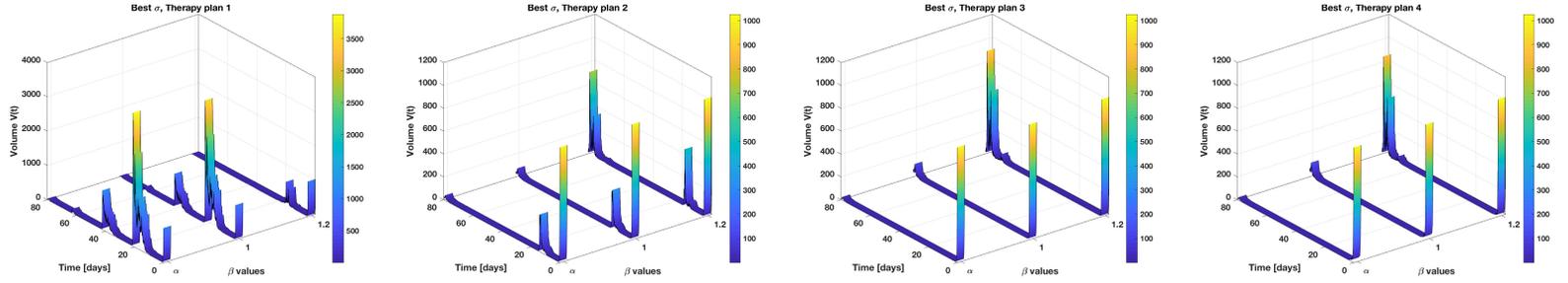

(c)

Fig. 6: **High fractal dimension (d = 2.98)**. (a) Comparison of different therapies for a specific stochastic multiplicative growth model, *"A1: DMTD, HDMTD and LDM better than MTD. DMTD, HDMTD and LDM are similar in term of recurrence but HDMTD and LDM have a better response in the transient phase (see zoom)."* (b) Effect of noise level on each therapy plan, *"A2: For MTD, DMTD, HDMTD and LDM, $\sigma^* = 0.5$ is better in terms of therapy effectiveness."* (c) Effect of $\beta$ for different plans and $\sigma^*$, *"A3: For MTD, $\beta = 1.2$ is better, and for DMTD, HDMTD and LDM, $\beta = \alpha$ is better".*

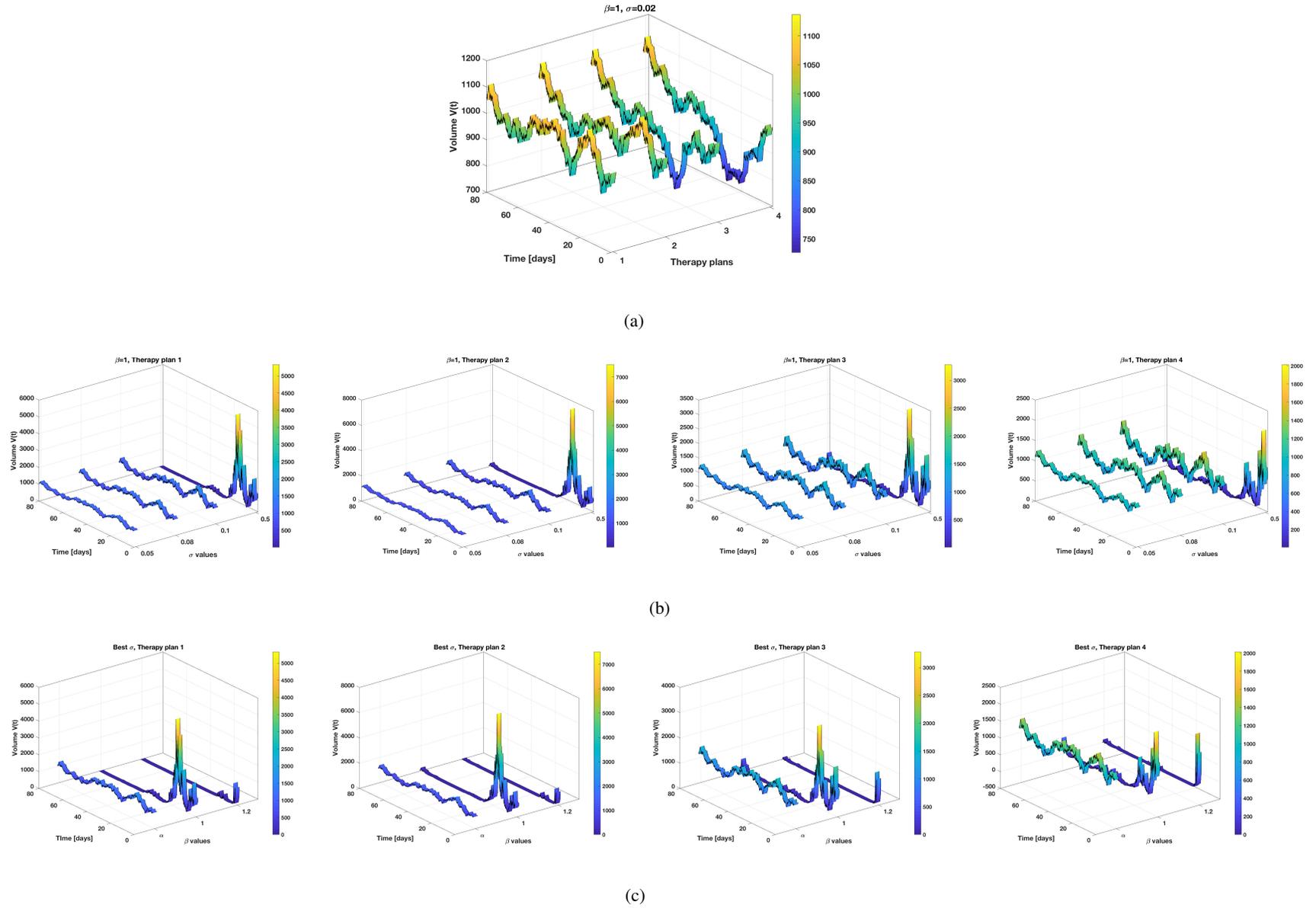

(a)

(b)

(c)

Fig. 7: **Low fractal dimension** (**d = 2.25**). (a) Comparison of different therapies for a specific stochastic multiplicative growth model, *"A1: HDMTD and LDM are better than MTD and DMTD."* (b) Effect of noise level on each therapy plan, *"A2: At the long run, σ* = 0.5 is better for all plans, even though the tumor increases at the start of therapy."* (c) Effect of β for different plans and σ*, *"A3: For all therapy plans, β = 1.2 is better in terms of effectiveness."*

where the state-space vector is $x(t) = [\bar{Z}(t), C(t)]^{tr}$, with $\bar{Z}(t) = \ln[V(t)]$, and

$$f(x,u) = \begin{bmatrix} \left(ae^{(\alpha-1)\bar{Z}} - b - \frac{\sigma^2}{2}\right)(1 - C(t)) \\ -kC(t) + u(t) \end{bmatrix}, \; g(\sigma) = \begin{bmatrix} \sigma & 0 \\ 0 & 0 \end{bmatrix}. \quad (31)$$

The values of the stochastic tumor growth model are the ones identified earlier using lung data and given in Table I. The elimination rate constant for the drug is $k = 1\,\mathrm{day}^{-1}$.

The objective is to find the optimal $u(t)$ that reduces the volume of cancer cells around a tolerable target volume value[6] $\bar{Z}^*$ while minimizing the effects of toxicity by constraining the amount of injected drug dose in $\mathscr{U}$. The optimal control problem to be solved is thus given as follows (see Figure 8 for a schematic diagram):

$$\begin{aligned} \min_u J(u) &:= \int_{t_k}^{t_k+T_p} q_1 \; (\bar{Z}(t) - \bar{Z}^*)^2 + q_2 \; u(t)^2 \, dt, \\ \text{s.t.} \; x(t_k) &= \hat{x}_{k|k}, \\ dx(t) &= f(x,u)\,dt, \\ u(t) &\in \mathscr{U} = [0, u_{\max}], \end{aligned} \quad (32)$$

where $T_p$ is the prediction horizon and the parameters $q_1$ and $q_2$ weight the relative importance of reducing the tumor volume and using less drug. The filtered state $\hat{x}_{k|k}$ is estimated using the EKF. As it can be seen from (32) and Figure 8, the design of the controller is constrained by a deterministic transformed NSM tumor growth model and tested on its stochastic counterpart. It is worth noting however that the solved OCP includes information about the noise level in the deterministic transformed dynamics, which is due to the applied Lamperti transformation.

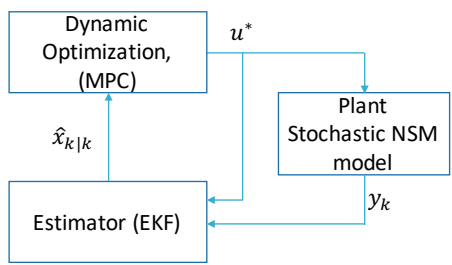

Fig. 8: Schematic diagram of the solved optimal control problem.

---

[6]In the sequel, $\bar{Z}(t)$ and $V(t)$ are used interchangeably in the text to refer to tumor volume.

The numerical solution of the OCP (32) is performed using the direct collocation method [31], [32]. The control input is assumed piecewise constant and parametrized as follows,

$$u(t) = \sum_{i=1}^{n} u_i \, \chi \left( [\tau_{i-1}, \tau_i[ \right), \tag{33}$$

where $\{\tau_i\}_{i=0}^{n}$, s.t., $\tau_0 = t_k$ and $\tau_n = t_k + T_p$, are the global collocation points, that define the local collocation points $\{\gamma_i\}_{i=0}^{m}$, s.t., $\gamma_0 = 0$ and $\gamma_m = 1$.

Clinical chemotherapy dosing differs from the proposed synthetic control systems technique in the manner of drug delivery. The designed control input is assumed to be piecewise constant, while the chemotherapeutic drug many times is not administrated continuously, but, e.g., as intravenous (IV) bolus. So, the piecewise designed controller is representative of the drug dosage and the values of the control at the collocation time points can be considered as the bolus injection dosage.

The simulation results were based on the Python code that was made available in "https://github.com/niclasbrok/nmpc_vdp". The code uses PyIpopt that allows to interface with Ipopt, which is an optimization software library, via Python. We chose $\bar{V}^* = 25\,\mathrm{mm}^3$, $q_1 = 100$, $q_2 = 1$ and $V(0) = 1000\,\mathrm{mm}^3 \sim 10^9$ cells corresponding to the size assumed to be clinically detectable, which were set in previous studies as well, e.g., in [12]. The prediction horizon $T_p = 10\,\mathrm{days}$, and the total simulation period is fixed to $140\,\mathrm{days}$. The allowable toxicity interval is chosen to be $\mathscr{U} = [0, 10]$.

Figures 9 and 10 show the optimal therapy schedules for the cases where volume measurements are available every $2\,\mathrm{days}$ and $7\,\mathrm{days}$ respectively. The global and local collocation points that define the subintervals of $u(t)$ are chosen as $n = 10$ and $m = 10$ respectively. The numerical results obtained suggest that an optimal chemotherapy plan that reduces the size of the tumor while minimizing the toxicity can be achieved by allowing the doses to vary during the treatment in response to the status of the tumor. This type of plans is not the standard of care in clinics yet but fits nicely to the type of adaptive therapy that is being a high field of investigation [33].

To assess the robustness of the proposed feedback control strategy to errors in the parameter values, we perturb the values of the parameters in the NSM model block while keeping the nominal values identified earlier for the dynamic optimization (MPC) and the estimator (EKF) blocks. The control parameters are kept the same as earlier, and the simulations were carried out for the case of 7 days interval between successive measurements.

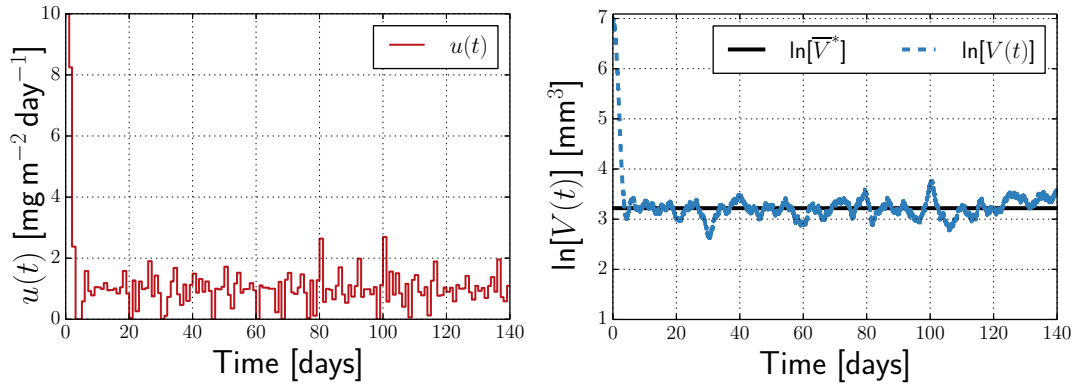

Fig. 9: **Volume measurements available every** $2$ **days**. Left panel: Optimal drug schedule. Right panel: Tumor volume.

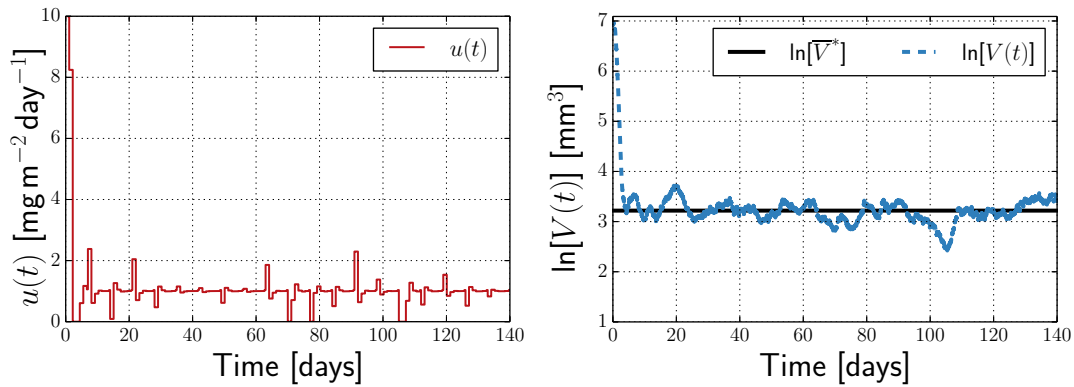

Fig. 10: **Volume measurements available every** $7$ **days**. Left panel: Optimal drug schedule. Right panel: Tumor volume.

Figures 11-13 show a certain degree of robustness against variations in growth parameters at the price of slower convergence. Moreover, the robustness is stronger when varying the growth and death parameters, $a$ and $b$, compared to the power law parameter, $\alpha$.

## VI. Conclusion and future research directions

In the present work, we investigated and analyzed a stochastic model describing the growth of cancerous cells at the macroscopic level. The model was extended from a mechanistic, deterministic model to account for the intrinsic and extrinsic sources of variability in the tumor growth process. We studied two variants of the proposed stochastic model, the unperturbed version where the growth is in the absence of a drug agent and the perturbed version in which the control drug is included.

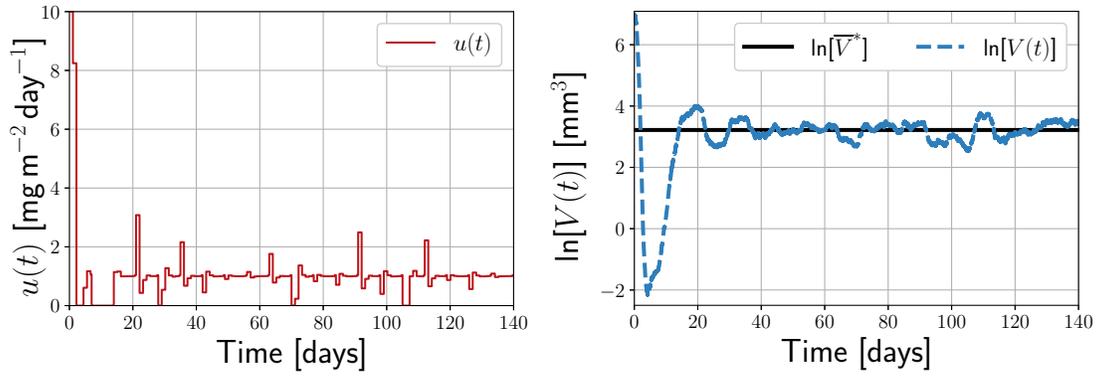

Fig. 11: Input and state profiles with 50 % variation in the growth parameter $a$

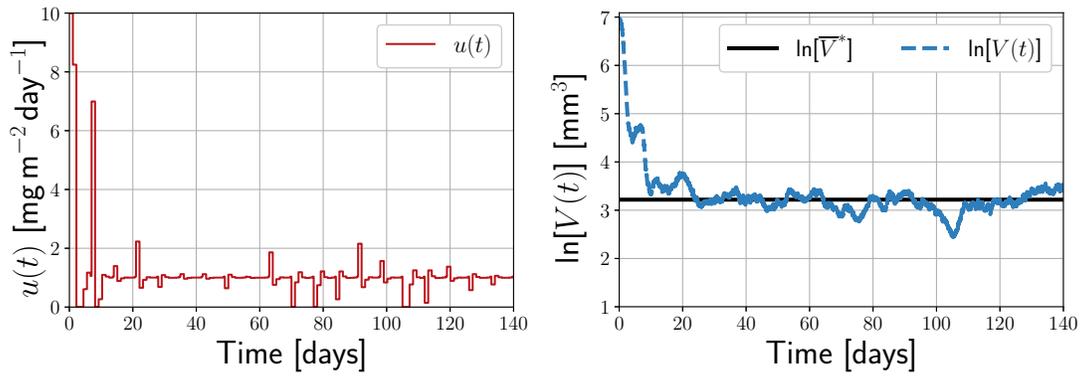

Fig. 12: Input and state profiles with 50 % variation in the death parameter $b$

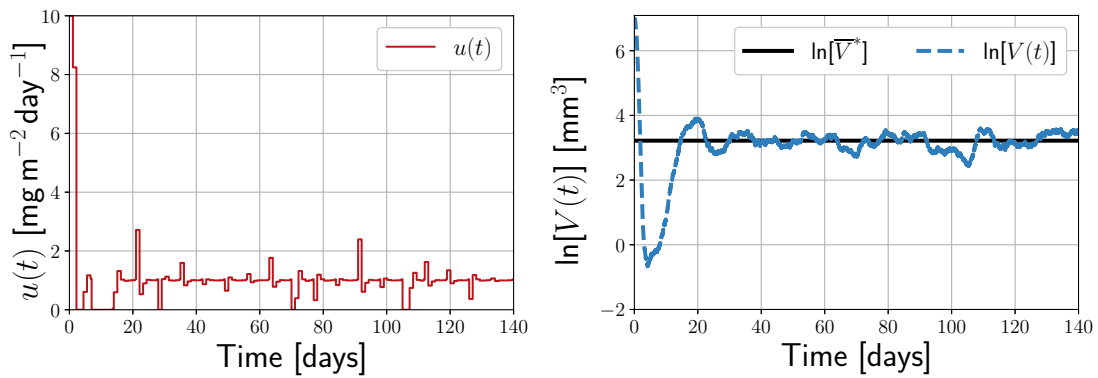

Fig. 13: Input and state profiles with 10 % variation in the power law parameter $\alpha$

We believe that the results we derived so far are promising in deriving some guiding principles that will hopefully allow for an increased understanding of the tumor growth process, as well as for designing improved therapy plans. Much work remains, and future investigation should be carried out to address a number of important issues towards possible clinical applications of the proposed stochastic uncontrolled/controlled tumor growth model.

First, we point out that the performance of the MLE-based estimation algorithm that is tested in Section IV is sensitive to the initial guess and thus may have converged to local optimum values. To circumvent this problem, we intend to use a Bayesian inference technique, similar to what was recently done in [34], to estimate the parameters of the stochastic NSM tumor growth model. We also plan to validate the identification approach using larger tumor volume datasets that will be collected at Memorial Sloan Kettering. We seek to find correlates between the parameter estimates for different types of cancer with an emphasis on the fractional power parameter $\alpha$ which is believed to be related to the density of tumor tissue. We will further analyze the predictive power of the identified stochastic growth model. Regarding the controlled model, we plan to solve the stochastic closed-loop OCP where probabilistic descriptions of uncertainties are incorporated into the OCP formulation. Instead of assuming that the noise term is multiplicative, the later extensions will be performed on the general stochastic NSM model where the diffusion term follows a law of allometry (i.e., $\beta$ will be considered as a degree of freedom for noise effect in addition to $\sigma$).

## Acknowledgements

This project was supported by AFOSR grant (FA9550-17-1-0435), grants from National Institutes of Health (R01-AG048769, R01-CA198121), MSK Cancer Center Support Grant/Core Grant (P30 CA008748), and a grant from Breast Cancer Research Foundation (grant BCRF-17-193).